\documentclass[a4paper,11pt]{article}
\usepackage{pdflscape}
\usepackage{amsmath}
\usepackage{cite}
\usepackage{amssymb}
\usepackage{dcolumn}
\usepackage{bm}
\usepackage{color}
\usepackage{epsfig}
\usepackage{amsfonts}
\usepackage{graphicx}
\usepackage{subfigure}
\usepackage{dcolumn}
\usepackage{indentfirst}
\usepackage{authblk}
\usepackage{slashed}
\usepackage{appendix}

\usepackage{booktabs}
\AtBeginDocument{
\heavyrulewidth=.08em
\lightrulewidth=.05em
\cmidrulewidth=.03em
\belowrulesep=.65ex
\belowbottomsep=0pt
\aboverulesep=.4ex
\abovetopsep=0pt
\cmidrulesep=\doublerulesep
\cmidrulekern=.5em
\defaultaddspace=.5em
}

\usepackage[colorlinks=false]{hyperref}
\hypersetup{citecolor=Blue}
\begin{document}

\vspace{80pt}

\centerline{\LARGE  Drag force on a moving heavy quark with deformed  string configuration }

\vspace{40pt}

\centerline{
Sara Tahery,$^{a}$ 
\footnote{E-mail: s.tahery@impcas.ac.cn}
Xurong Chen $^{b}$ 
\footnote{E-mail: xchen@impcas.ac.cn}
}
\vspace{30pt}

{\centerline {$^{a,b}${\it Institute of Modern Physics, Chinese Academy of Sciences, Lanzhou 730000,  China
}}
\vspace{4pt}
{\centerline {$^{a,b}${\it University of Chinese Academy of Sciences, Beijing 100049, China
}} 
\vspace{4pt}
{\centerline {$^{b}${\it  Guangdong Provincial Key Laboratory of Nuclear Science, Institute of Quantum Matter,}}
{\centerline {$^{}${\it  South China Normal University, Guangzhou 510006, China}}

 \vspace{40pt}

\begin{abstract}
To study drag force on a moving heavy quark through a plasma, we use a deformed AdS space-time, in which deformation parameter $c$  describes non-conformality in AdS/QCD. In this case the quark is mapped to a probe string in the AdS space. Considering probable contribution of deformation parameter in the probe string, we apply a general form of  c-dependent string ansatz in the drag force computation. Then we find the acceptable value of this parameter as it satisfies QCD calculations. Using this result, we also discuss  diffusion constant which is in agreement with phenomenological result for non-relativistic limit. Also we show that while in absence of deformation parameter, probe string is a strictly increasing function of radial coordinate, the c-dependent probe string  has a maximum value versus $z$.
\end{abstract}

\newpage

\tableofcontents

\newpage

\section{Introduction}
Studying motion of quarks through QGP\footnote{Quark Gluon Plasma} during heavy ion collision  is a distinctive feature of RHIC\footnote{Relativistic Heavy Ion Collider} data. In fact when a high energy parton propagates through the QGP, it quenches strongly. The energy loss is
 analyzed by drag force, which is related to the interaction between the moving quark and the medium. There is a connection between string theory and relativistic heavy
ion collisions. The energy loss of heavy quark is
understood as the momentum flow along a moving classical  string into the horizon. Perturbative calculation of many QGP quantities are currently out of our reach in the strong-coupling regime but gauge/gravity  correspondence  has been used to investigate observable quantities in various interesting strongly-coupled gauge theories where perturbation theory is not applicable.\\
AdS/CFT\footnote{Anti de sitter Space-time/ Conformal Field Theory}  conjecture originally relates the type IIB string theory on $AdS_5\times S_5$
space-time to the four-dimensional $N=4$ SYM\footnote{Super-Yang-Mills} gauge theory \cite{adscft}. Also, many other studies 
represent a holographic description of AdS/CFT in which, a strongly coupled field theory 
on the boundary of the AdS space is mapped to the weakly coupled gravity theory in
the bulk of AdS \cite{holo}.

In this conjecture  the motion of a quark through a plasma is described by a mechanism in which there is
a simple holographic picture dual to the probe quark moving in
plasma, as the quark is mapped to a probe string in the AdS space. So instead of studying quark's motion in a strongly coupled system one can simply consider the classical dynamics of a string in gravity side.\\
 In AdS/QCD correspondence replacing QCD\footnote{Quantum Chromo Dynamics} by $N = 4$ SYM theory is an approximation according to the fact that after a critical temperature $N = 4$ SYM theory  may capture much of the dynamics
of the QGP  so that confinement and the chiral
condensate have disappeared. This temperature is high enough to do this  but low enough so that the ’t Hooft coupling is still large. Therefore when a high-energy parton passes through the QGP, one may consider an external quark,
in the framework of AdS/CFT \cite{gath}. The drag force formalism in AdS/CFT has been studied in \cite{drag,enlo} to compute the energy loss, later the obtained method was applied in \cite{codr}. In the mentioned papers the
momentum rate flowing down to the  probe string has been interpreted as drag force exerted
from the plasma on the quark. Where the quark is prescribed to move on the boundary of $AdS_5$. For the heavy quark the most sensible trajectories
are those with constant velocity relative to the reference frame, rest frame of the plasma. The string trails out behind the quark and dangles into $AdS_5$. It is a holographic representation of the color flux from the external quark spreading out in the 3 + 1 dimensions of the boundary theory.

AdS/QCD, is an approach in which one starts from a five-dimensional effective
field theory somehow motivated by string theory and tries to fit it to QCD as much as possible. Since $AdS_5$ is the gravity dual of $N = 4$ SYM that is an approximation of QCD, it would be interesting to consider some corrections in $AdS_5$ as the gravity dual to find more phenomenological results. Motion of a quark  has been studied in different literature \cite{neme,acen,rost,caje,brmo,stst,nore,inth,hequ,en,drma} by AdS/QCD. In \cite{corr} the second correction of radial coordinate has been discussed where the coefficient of such a correction is called as deformation parameter. Different effects of such a parameter on physics of quark and it's motion are studied in \cite{qupo,dfsp,vac}. Also  the effect of deformation parameter on drag force on di-quark is considered in \cite{dfga}.
During recent years, different computations of drag force have been done \cite{dch,dge,dsy,dget,stmo,scl,r2,chef,thtr,hth,nonre,anis,anpl,anch,chair,dins,heq,vacu,
alor,hQCD}.

 In this paper we will study the effect of deformation parameter while it contributes in the  probe string ansatz as well. In other words we test if with a deformed AdS background, the string ansatz could be affected with an extra term representing deformation parameter. So we will define a general deformed  probe string  to compute the drag force.\\
This paper is organized as follows, in section (\ref {se:drag force}) we will review  holographic description of drag force, then in section (\ref{sec:def}) temporarily we  will quit the drag force discussion and  skip to review of deformed AdS. After introducing the method and the background both, we will compute the drag force in a deformed AdS in section (\ref {dfdeformed}). According to the obtained theoretical results we will discuss some phenomenological aspects  of the system in section (\ref{se:ph}). We will end up by conclusion in section (\ref{se:con}).
\section{Review of holographic description of drag force}\label{se:drag force}
This section is a review of  drag force computation according to AdS/CFT approach based on \cite{drag}. So the reader who is familiar with this reference may skip this section.\\
Let us consider a moving external quark in a thermal plasma. In frame work of AdS/QCD  such a quark is considered as a  probe string dangling from the boundary of the field theory where the end point of the  probe string describes the quark in one higher dimensional space-time dual to the field theory.\\
In general, one can consider a  probe string which is described in the AdS by NG\footnote{Nambu-Goto} action  as,
\begin{equation}\label{eq:NG action}
S=-\frac{1}{2\pi\alpha'} \int d^2 \sigma \sqrt{-\det  g_{\alpha\beta} }\quad\quad g_{\alpha\beta}=G_{\mu \nu} \partial X_{\alpha}^{\mu} \partial X_{\beta}^{\nu},
\end{equation}
where $\sigma^{\alpha}$ are coordinates  of the string worldsheet and embedding of the string worldsheet in space-time is specified as,
\begin{equation}\label{eq:X}
X^{\mu}=X^{\mu}(\sigma)=(X_0,X_1,X_2,X_3,X_4),
\end{equation}
also $G_{\mu\nu}$ is the five-dimensional Einstein metric.\\
The equation of motion   is obtained from \eqref{eq:NG action} as,
\begin{equation}\label{eq:eqmo}
\nabla_{\alpha} P^{\alpha}_{\mu}=0\quad\quad P^{\alpha}_{\mu}=-\frac{1}{2\pi \alpha'} G_{\mu \nu} \partial^{\alpha} X^{\nu},
\end{equation}
where $\nabla_{\alpha}$ is the covariant derivative with respect to $g_{\alpha\beta}$.
 $P^{\alpha}_{\mu}$ is the worldsheet current of
space-time energy momentum carried by the  probe string. Recall that $\mu$ is space-time index and $\alpha$ index stands for worldsheet coordinate.  Considering  a moving quark  along arbitrary direction  as $X_i$, to calculate the flow of momentum $\frac{dp_i}{dt}$ down the string, one needs the following integral,
\begin{equation}\label{eq:deltapone}
\Delta P_i=\int_{\mathcal{I}} dt \sqrt{-g} P^{r}_{x^i}=\frac{dp_i}{dt}\Delta t,
\end{equation}
where the integral is taken over some time interval $\mathcal{I}$ of length $\Delta t$.
 Drag force should be in opposite direction of the motion, so the $\frac{dp_i}{dt}$ is a negative quantity. After some calculation the drag force is obtained as,
\begin{equation}\label{drag force}
\frac{dp_i}{dt}=\sqrt{-g}  P^{r}_{x^i}= -\frac{1}{2\pi\alpha'} \pi_{i},
\end{equation}
where $\pi_{i}$ is the  conjugate momentum $\pi_{i}=\frac{\partial\mathcal{L}}{\partial X ^{\prime}_i}$.\\
Also the rate of the energy loss for a moving quark with speed $v$ is, 
\begin{equation}\label{energy loss}
\frac{dE}{dt}=\sqrt{-g}  P^{r}_{t}= -\frac{1}{2\pi\alpha'} \pi_{t}=-\frac{v}{2\pi\alpha'} \pi_{i},
\end{equation}
where $\pi_{t}=\frac{\partial\mathcal{L}}{\partial \dot{X}}$.\\
In continue we consider a general  background metric as, 
\begin{equation}\label{eq:metric}
ds^2= G_{00} dt^2+ G_{ii}\Sigma_{i=1}^{i=3} dx^2_{i}+ G_{zz} dz^2,
\end{equation} 
where the dynamic of the  probe string is described by NG action.
If the motion of the string is considered in one direction as $X_3(t,z)$ then,
\begin{equation}\label{eq:Xsigma}
X^{\mu}(\sigma)=(t,X_1,X_2,X_3(t,z),z). 
\end{equation} 
 From \eqref{eq:NG action} and \eqref{eq:X} we will have, 
\begin{equation}\label{eq:lag}
S=\frac{1}{2\pi\alpha'} \int dt dz \mathcal{L}\quad  \quad\quad \quad \quad  \mathcal{L}=-\sqrt{-\det g_{\alpha\beta} }.
\end{equation}
From \eqref{eq:Xsigma} we write,
\begin{eqnarray}\label{eqn:g}
g_{tt}&=&G_{\mu \nu} \partial_{t} X^{\mu} \partial_{t} X^{\nu}\nonumber\\
g_{tz}&=&G_{\mu \nu} \partial_{t} X^{\mu} \partial_{z} X^{\nu}\nonumber\\
g_{zt}&=&G_{\mu \nu} \partial_{z} X^{\mu} \partial_{t} X^{\nu}\nonumber\\
g_{zz}&=&G_{\mu \nu} \partial_{z} X^{\mu} \partial_{z} X^{\nu},\nonumber\\
\end{eqnarray}
Considering $G_{\mu \nu}$ from the background metric \eqref{eq:metric} and $X^{\mu}(\sigma)$  from  \eqref{eq:Xsigma}  the relations \eqref{eqn:g} lead to,
\begin{eqnarray}\label{eqn:gg}
g_{tt}&=&G_{00}+G_{33}\dot{X_{3}}^{2}\nonumber\\
g_{tz}&=&G_{33}\dot{X_{3}} X_{3}'\nonumber\\
g_{zt}&=&G_{33} X_{3}'\dot{X_{3}}\nonumber\\
g_{zz}&=&G_{33}X_{3}'^{2}+G_{zz},\nonumber\\
\end{eqnarray} 
so from \eqref{eq:lag} and \eqref{eqn:gg} the lagrangian is derived as,
\begin{equation}\label{eq:lagmetric}
\mathcal{L}=-\sqrt{-[G_{00}G_{33}X_{3}'^2+G_{zz}G_{33} \dot{X_{3}}^2+G_{00}G_{zz}]},
\end{equation}
where $\dot{X_{i}}$ and $X'_{i}$ stand for $\frac{\partial X_{i}}{\partial t}$ and  $\frac{\partial {X_{i}}}{\partial z}$ respectively.\\
The conservation equation of conjugate momentum implies,
\begin{equation}\label{eq:momentum}
\pi_{3}=\frac{G_{00}G_{33} X'_{3}}{\sqrt{-[G_{00}G_{33} {X ^{\prime}_3}^{2}+G_{zz}G_{33} \dot{X_{3}}^2+G_{00}G_{zz}]}}.
\end{equation}
By solving \eqref{eq:momentum} for $X ^{\prime}_3$ we obtain,
\begin{equation}\label{Xprim}
X ^{\prime}_3=\pm \pi_{3}\sqrt{\frac{-G_{zz}}{(G_{00}G_{33}+\pi^2_{3})}(\frac{\dot{X_{3}}^2}{G_{00}}+\frac{1}{G_{33}})}.
\end{equation}
The profile $X_3$ should be describing a string that trails out behind the quark, not in front of it, so the  sign behind $\pi_{3}$ in (\ref{Xprim}) should be chosen positive.\\
Before integrating $X ^{\prime}_3$ to find $X_3$ we must require a way to avoid an imaginary right hand side in \eqref{Xprim}. The only way is to impose that is, the numerator and denominator have common root, means both of them should be zero at a specific $z$. Then, the result of such a computation from \eqref{Xprim} should be applied in \eqref{eq:momentum} to obtain the well defined conjugate momentum, and then drag force from \eqref{drag force}. 
Having reviewed the drag force computation, we quit more details of that yet and in continue will discuss a deformed AdS background in section {\ref{sec:def}}. 
\section{Review of deformed asymptotically AdS background}\label{sec:def}
In this section we review motivation of considering the second correction of radial coordinate based on references \cite{corr,qupo}.\\
 AdS/CFT correspondence is applied in strong interactions. It should be done by finding string description. Recall that $AdS_5$ is not an exact dual of QCD  but of $N=4$ SYM. In fact, QCD is not a conformal theory. So the approach to find it's string description is to start from a five-dimensional effective field theory  motivated by string theory and try to fit it to QCD as much as possible.

Although the original conjecture of AdS/CFT was for conformal theories, but some modifications produce mass gap, confinement and supersymmetry breaking in gauge/gravity duality. In this way the produced model shares a few key features with QCD
that makes it more useful for phenomenology than $AdS_5$.
In AdS/QCD approach one may address the issue of the quadratic correction within the simplified models. It is useful to adopt the geometric approach and estimate the correction and compare the result
with that of QCD.\\
The idea of modifying pure AdS geometry has been discussed in \cite{geap} in detail.  Here we explain some headlines of motivation and approach. The interested reader is recommended to follow the calculation and detail in the main reference. The idea comes from the comparison of high energy correlators with OPE of two point functions of QCD quark current. The question is,\textit{ how does the 5D dual of QCD look like?} In \cite{moho,adfl,toho,mesp,chsy,chdy,flgr,loen,hasp} bottom-up approach succeeded in the study of the aspects the dual should explain, independently of a stringy set-up. The point is, QCD models which are defined on truncated AdS space, lack power corrections in the correlator. Therefore to incorporate more feature of QCD, one may consider showing non-conformality in the metric. In \cite{geap} modification of metric has been done in comparison with OPE, where the logarithm part of OPE has been produced by considering $\omega(z)=\frac{l_0}{z}$ in  a purely AdS metric in
conformal coordinates and string frame as,
\begin{equation}\label{nonconformal metric}
ds^2=\omega^2(z)(\eta_{ab} dx^a dx^b-dz^2),
\end{equation}
and the power corrections in OPE can be generated by deviations from conformality with a given power of $z^{2d}$ in the metric. Later in \cite{corr,qupo} author estimated the quadratic correction of such a metric background and fixed the coefficient of that. It was done by considering meson operators and looking for the corresponding vertex operators. The detail of those references and all steps and calculations are beyond of the goals of our study and we will take advantage of the found metric. The interested reader can consider them. Briefly, from those studies we find that in particular one can consider  a rather non-trivial structure of the quark configuration for the warp factor in the background metric. From asymptotic linearity of Regge trajectories one may understand that some backgrounds reduce to the standard AdS background in the UV but differ from it in the IR. It is known that  some strong interactions would include the dominant square 
term at short distances as well as the dominant linear term at large distances. This assumption is quite acceptable in classical  limit of string theory where we are working. In this way one can find  aspects of different complicated warp and blackening  factors.\\ 
We  consider the following background metric (in the string frame),
\begin{equation}\label{eq: metric}
ds^2=\frac{R^2}{z^2} h(z)  (-f(z) dt^2+\Sigma_{i=1}^{3}dx_i^2+\frac{1}{f(z)} dz^2),
\end{equation}
where \footnote{The sign of the exponent in (\ref{eq:metricT}) is different from that of
\cite{corr}, because in the mentioned reference the author used the Euclidean metric.
The Minkowskian metric of \cite{corr}  is obtained via analytic
continuation $x \longrightarrow -ix$ together with $z\longrightarrow -iz$.},
\begin{equation}\label{eq:metricT}
f(z)=1-\left(\frac{z}{z_h}\right)^4,\quad\quad h(z)=e^{-\frac{c}{2}z^2 },
\end{equation}
the horizon is located at $z=z_h$  and the temperature of the black hole is $T=\frac{1}{\pi z_h}$. $c$ stands for the coefficient of second correction of radial coordinate of the background, called the deformation parameter. Note  at  limit $c\longrightarrow 0$, the $AdS_5$ metric will be obtained. As $c$ becomes more than zero, it deforms the AdS space. This model has some phenomenological benefits as, it is a nearly conformal theory at UV, it results in linear Regge-like spectra for mesons \cite{corr,lincon} and results in a phenomenologically satisfactory description of the confining potential \cite{qupo}.
In the next section we will use the deformed background.
\section{Drag force on a moving heavy quark in a plasma with deformed gravity dual } {\label{dfdeformed}}
Let us study  the drag force in the deformed $AdS_5$ (\ref{eq: metric})  while the configuration of the  probe string attached to the moving quark  also is affected by deformation parameter. Firstly  we remind  an example of another background  to clarify the motivation of our case of interest.  For studying drag force in presence of magnetic field, one may consider a magnetized metric which leads to some  effects on the string ansatz as well. So, in presence of deformation parameter of metric, the question arises as whether such a parameter appears in string configuration? To find it, we need a  suitable ansatz to describe the behavior of the  probe string attached to the moving quark. The massive heavy quark is prescribed to move along trajectories with constant velocity relative to the reference frame (the rest frame of the plasma). The ansatz should satisfy the assumption that at late time the steady state behavior is obtained. With the aim of  phenomenological verification of  presence of parameter $c$ in the string ansatz  we define it with a probable contribution of deformation parameter in a general form as follows, 
\begin{eqnarray}\label{eq:x3}
X_0&=&t,\nonumber\\
X_1&=&0,\nonumber\\
X_2&=&0,\nonumber\\
X_3(z,t)&=&vt+\varepsilon(z)+ \mathcal{O}(t),\quad\quad \varepsilon(z)=\zeta_0(z)+({\frac{c}{T^2}}) \zeta(z), \nonumber\\
X_4&=&z,
\end{eqnarray}
where $c$ is the deformation parameter  in (\ref{eq: metric}) and (\ref{eq:metricT}), so $\frac{c}{T^2}$ is a dimensionless variable. Also
 $\mathcal{O}(t)$ are all terms which vanish at late time, from now on we ignore them.
From (\ref{eq:x3}) we have, 
\begin{eqnarray}\label{xdotxprim}
\dot{X_3}&=&v,\nonumber\\
X'_3&=&\varepsilon'(z)\nonumber\\
\varepsilon'(z)&=& \zeta_0'(z)+ ({\frac{c}{T^2}}) \zeta'(z),
\end{eqnarray} 
where $'$ stands for derivative with respect to $z$.\\
Now, with (\ref{eq: metric}), (\ref{eq:metricT}) and (\ref{eq:x3}) the relation (\ref{eq:lagmetric}) is given as ,
\begin{equation}\label{eq:Lagrangy}
\mathcal{L}=-\frac{R^2 }{z^2}e^{-\frac{c}{2}z^2 }\sqrt{1+f(z)( \zeta'_0(z) + ({\frac{c}{T^2}}) \zeta'(z) )^2- \frac{v^2}{f(z)}}.
\end{equation}
Then from \eqref{eq:momentum} and (\ref{xdotxprim}) the conjugate momentum is written as,
\begin{eqnarray}\label{eq:pi}
\pi_{3}&=&\frac{\partial \mathcal{L}}{\partial X_3'}=\frac{\partial \mathcal{L}}{\partial \varepsilon'(z)}=\frac{\partial \mathcal{L}}{\partial ( \zeta'_0(z) + ({\frac{c}{T^2}}) \zeta'(z) )}\nonumber\\
&=&\frac{\partial \mathcal{L}}{\partial \zeta'_0(z)+ ({\frac{c}{T^2}}) \partial \zeta'(z)}\nonumber\\
&=&\frac{1}{\frac{1}{\frac{\partial \mathcal{L}}{\partial \zeta'_0(z)}}+ ({\frac{c}{T^2}}) \frac{1}{\frac{\partial \mathcal{L}}{\partial \zeta'(z)}}}.
\end{eqnarray}
Using (\ref {eq:Lagrangy}) results in, 
\begin{eqnarray}\label{eq:allpi}
\frac{\partial \mathcal{L}}{\partial \zeta'_0(z)}&=& -\frac{ R^2 }{z^2} e^{-\frac{c}{2}z^2 } \frac{ f(z)( \zeta'_0(z) + ({\frac{c}{T^2}}) \zeta'(z) )}{\sqrt{1+f(z)( \zeta'_0(z) + ({\frac{c}{T^2}}) \zeta'(z)  )^2- \frac{v^2}{f(z)}}},\nonumber\\
  \frac{\partial \mathcal{L}}{\partial \zeta'(z)}&=& -\frac{ R^2 }{z^2} e^{-\frac{c}{2}z^2 } \frac{({\frac{c}{T^2}}) f(z)( \zeta'_0(z) + ({\frac{c}{T^2}}) \zeta'(z) )}{\sqrt{1+f(z)( \zeta'_0(z) + ({\frac{c}{T^2}}) \zeta'(z)  )^2- \frac{v^2}{f(z)}}},
\end{eqnarray}
plugging \eqref{eq:allpi} in (\ref{eq:pi}) leads to,
\begin{equation}\label{eq:pinew}
\pi_{3}=-\frac{ R^2 }{z^2} e^{-\frac{c}{2}z^2 } \frac{ f(z)( \zeta'_0(z) + ({\frac{c}{T^2}}) \zeta'(z) )}{2\sqrt{1+f(z)( \zeta'_0(z) + (\frac{c}{T^2}) \zeta'(z) )^2- \frac{v^2}{f(z)}}}.
\end{equation}
Proceeding by  solving (\ref{eq:pinew}) for the string ansatz we obtain,
\begin{equation}\label{eq:epsilonprim}
 \zeta'_0(z) + ({\frac{c}{T^2}}) \zeta'(z) =\pm \pi_{3} \sqrt{\frac{1-\frac{v^2}{f(z)}}{\frac{ R^4 }{4z^4}e^{-cz^2}f^2(z)-f(z)\pi^2_{3}}},
\end{equation}
since the string trails out behind the quark the acceptable sign in (\ref{eq:epsilonprim}) is +.
To avoid imaginary string ansatz, the right hand side of (\ref{eq:epsilonprim}) should be real, means the nominator and denominator should have common root. By applying this condition, necessarily the final result for momentum is, 
\begin{equation}\label{eq:pisq}
\pi_{3}=\frac{R^2 }{2z_h^2} \frac{v}{\sqrt{1-v^2}} e^{-\frac{c}{2}z_h^2 \sqrt{1-v^2}}.  
\end{equation}
where $R^2=\sqrt{N}\alpha' g_{YM}$ and $z_h=\frac{1}{\pi T}$, T is Hawking temperature and it's dual description is temperature of the plasma. Then from (\ref{drag force}) and \eqref{eq:pisq} the final result for the drag force is
\begin{equation}\label{eq:Fv}
\frac{dp_3}{dt}=-\frac{\pi}{4}\sqrt{N} g_{_{YM}} T^2 \frac{v}{\sqrt{1-v^2}} e^{-\frac{c}{2 \pi^2 T^2} \sqrt{1-v^2}}.
\end{equation}
and the rate of the energy loss from (\ref{energy loss}) is written as,
\begin{equation}\label{eq:Ev}
\frac{dE}{dt}=-\frac{\pi}{4}\sqrt{N} g_{_{YM}} T^2 \frac{v^2}{\sqrt{1-v^2}} e^{-\frac{c}{2\pi^2 T^2} \sqrt{1-v^2}}.
\end{equation}
Also (\ref{eq:pisq}) with (\ref{eq:epsilonprim}) together lead to,
\begin{equation}\label{eq:zetaandzetaprim}
\zeta'_0(z) + ({\frac{c}{T^2}})\zeta'(z)= \frac{1 }{z_h^2} \frac{v}{\sqrt{1-v^2}}  \sqrt{\frac{(1-\frac{v^2}{(1-(\frac{z}{z_h})^4)})}{(1-(\frac{z}{z_h})^4)[(\frac{1}{z^4}-\frac{1}{z_h^4})e^{-c(z^2-z_h^2 \sqrt{1-v^2})}-\frac{1 }{z_h^4} \frac{v^2}{1-v^2}]}}.
\end{equation}
Since in \eqref{eq:zetaandzetaprim}, $\zeta'_0(z)$ is derivative of string configuration at limit $c\longrightarrow 0$, using this condition  we may write,
\begin{equation}\label{eq:zetaprim}
\zeta'_0(z)=\frac{v z^2 z_h^2}{z_h^4-z^4},
\end{equation}
which leads to,
\begin{equation}
\zeta_0(z)=-\frac{v}{2} z_h[\tan^{-1}\frac{z}{z_h}+\ln\sqrt{\frac{z_h-z}{z_h+z}}].
\end{equation}
Later we will use the differential equation  (\ref{eq:zetaandzetaprim}) to find the second term in right hand side of (\ref{eq:x3}).\\
We will continue by discussing the value of  parameter $c$ regards to QCD in the next section.
\section{ Discussions} \label{se:ph}
 Since the drag force is in $X_3$ direction, from now on we remove index $3$ for simplicity. To find the deformation parameter we need to solve the equation (\ref{eq:Fv}), in the limit of large $N$ and large $Ng^2_{_{YM}}$  we have $\frac{p}{m}=\frac{v}{\sqrt{1-v^2}}$.  So from (\ref{eq:Fv}) we find,\footnote{Although our gravity dual is not conformal, we  use information from the conformal system. The reason is, firstly, QCD is not a conformal theory itself. Needless to remind that AdS/QCD is not an exact AdS/CFT duality but a very good approximation of that. Therefore even with $c=0$, one uses  the limit of large $N$ and large $Ng^2_{_{YM}}$  for QCD which is not conformal! Considering a non-conformal metric with $c\neq 0$ is nothing but describing  QCD with a better gravity dual than $AdS_5$. So still the  data on the  boundary theory which should be matched has not been changed  phenomenologically. }
\begin{equation}\label{eq:Fp}
\frac{dp}{dt}=-\frac{\pi}{4}\sqrt{N}g_{_{YM}} T^2 \frac{p}{m} e^{-\frac{c}{2\pi^2 T^2} \frac{m}{\sqrt{p^2+m^2}}},  
\end{equation}
 and the momentum of a moving heavy quark in a  plasma falls by $\frac{1}{e}$ in a time \cite{codr},
\begin{equation}\label{relaxation time}
t_0=\frac{2m}{\pi \sqrt{N}g_{_{YM}} T^2},
\end{equation}
or
\begin{equation}\label{eq:pandt0}
p(t)=p_1(0)e^{-\frac{t}{t_0}}.
\end{equation}
After plugging (\ref{eq:pandt0}) into left hand side of (\ref{eq:Fp}), and by using (\ref{relaxation time}) we  integrate both sides from $t=0$ to $t\rightarrow \infty$. So the relation between temperature of the QGP and deformation parameter on gravity side is given as,
\begin{eqnarray}\label{eq:c}
c&=&-2\pi^2T^2 \ln\frac{4m}{t_0\pi\sqrt{N}g_{_{YM}}T^2}= -2\pi^2T^2 \ln 2. 
\end{eqnarray} 
From above we write the exponential in \eqref{eq:Fp} as,
\begin{equation}\label{expo}
 e^{-\frac{c}{2\pi^2 T^2} \sqrt{1-v^2}}= 2^{\sqrt{1-v^2}}.
\end{equation}
Plugging (\ref{eq:c}) and (\ref{expo}) in (\ref{eq:Fv}) and (\ref{eq:Ev}) both,  the drag force and the energy loss are written as follows,
\begin{equation}\label{eq:Fv2}
\frac{dp}{dt}=-\frac{\pi}{4}\sqrt{N} g_{_{YM}} T^2 \frac{v}{\sqrt{1-v^2}} 2^{\sqrt{1-v^2}},
\end{equation}
\begin{equation}\label{eq:Ev2}
\frac{dE}{dt}=-\frac{\pi}{4}\sqrt{N} g_{_{YM}} T^2 \frac{v^2}{\sqrt{1-v^2}}  2^{\sqrt{1-v^2}}.
\end{equation}
 Recall that function $h(z^2), h(0)=0$ in \eqref{eq:metricT} defines the deformation parameter in Euclidean space or Minkowskian space (see \cite{corr,qupo,heq,deformation} to compare $c$). In fact $c^2$ appears in energy momentum tensor \cite{term}. So phenomenologically, the magnitude of $c$ in \eqref{eq:c} is a comparative  parameter with  QCD data.
At the temperature $T=250$ MeV of QCD \cite{codr}, deformation parameter  in (\ref{eq:c}) is $\vert c \vert\simeq 0.84$ Ge$V^2$  which is in agreement with the result of \cite{corr}.

Another quantity that could be studied is  diffusion constant which is defined by \cite{codr},
\begin{equation}\label{D}
D=\frac{T}{m} t_D, 
\end{equation}
where $T$ is temperature, $m$ is heavy quark mass and $t_D$ is damping time which is related to drag force as follows\footnote{There are differences between the current calculation and what we have done with (\ref{relaxation time}) and (\ref{eq:pandt0}) relations. In fact, to solve an equation we considered  both relaxation time and momentum of heavy quark in SYM as  conditions. As an interesting practice/check, one can compute $t_D$ in $SW_{T,\mu} $  model  \cite{heq} and compare that result at limit $Q\longrightarrow 0$ with our result,  then by considering  value of $c$ may find the agreement.},
\begin{equation}\label{tD}
f \int _{0}^{t_D}dt=\int_{p}^{0}dp, 
\end{equation}
therefore (\ref{eq:Fv2}) is written as,
\begin{equation}\label{Dfin}
D=\frac{4}{\pi \sqrt{N}g_{_{YM}} T 2^{^{\sqrt{1-v^2}}}},
\end{equation}
 and by considering $\sqrt{N}g_{_{YM}}=\sqrt{5.5} $ as the best fit to QCD \cite{codr} we obtain,
\begin{equation}\label{Dvalue}
D=\frac{0.54}{T}\frac{1}{ 2^{\sqrt{1-v^2}}},
\end{equation}
as the  diffusion constant of a  heavy moving quark in a QGP with a deformed background.\\
For heavy quark, at non-relativistic limit $v^2\ll 1$ ( in natural unit), the result of \eqref{Dvalue} is close to the phenomenological result of \cite{diffu}. \\
\begin{figure}[h!]
\begin{center}$
\begin{array}{cccc}
\includegraphics[width=100 mm]{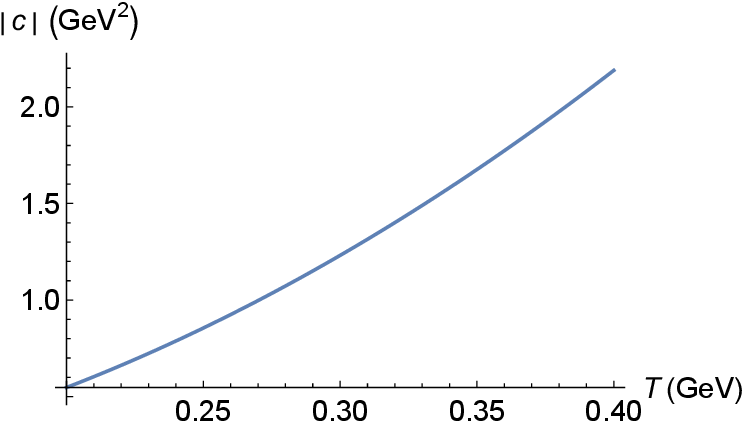}
\end{array}$
\end{center}
\caption{Behavior of deformation parameter with respect to temperature, from \eqref{eq:c}.}
\label{cT.eps}
\end{figure}
 Fig.~\ref{cT.eps} shows the behavior of magnitude of deformation parameter  with respect to the temperature of the plasma. The plot shows increasing temperature increases deformation parameter.
\begin{figure}[h!]
\begin{center}$
\begin{array}{cccc}
\includegraphics[width=100 mm]{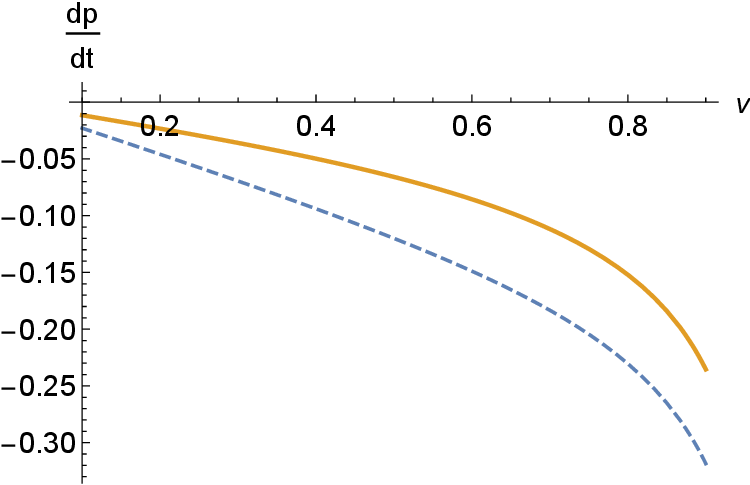}
\end{array}$
\end{center}
\caption{Drag force with respect to speed. Orange thick plot shows the drag force in absence of deformation parameter, or $c\longrightarrow 0$ in (\ref{eq:Fv}) and dashed blue plot shows it in presence of deformation parameter.  The speed of light is taken as unit, $T=250 MeV$ and $N g^2_{_{YM}}=5.5$ \cite{codr}.}
\label{fv.eps}
\end{figure}
In fig.~\ref{fv.eps} we study drag force versus  speed of moving quark. Magnitude of the drag force increases with increasing speed of moving quark, also nonzero values of  deformation parameter strengthen  the drag force. 
\begin{figure}[h!]
\begin{minipage}[c]{1\textwidth}
\tiny{(a)}\includegraphics[width=7cm,height=6cm,clip]{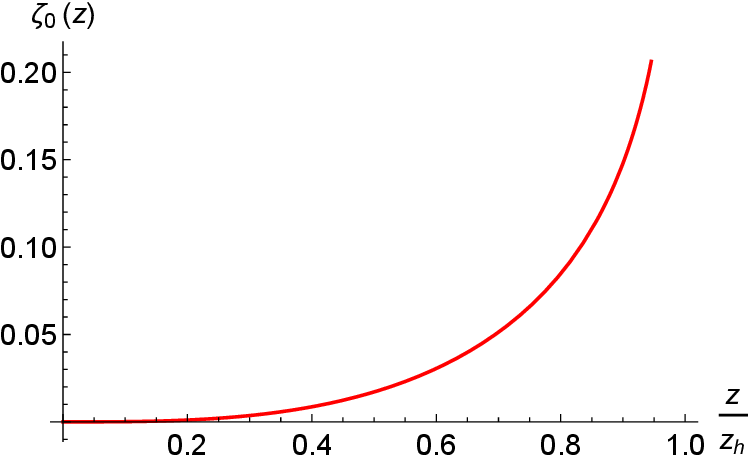}
\hspace{0.3cm}
\tiny{(b)}\includegraphics[width=7cm,height=6cm,clip]{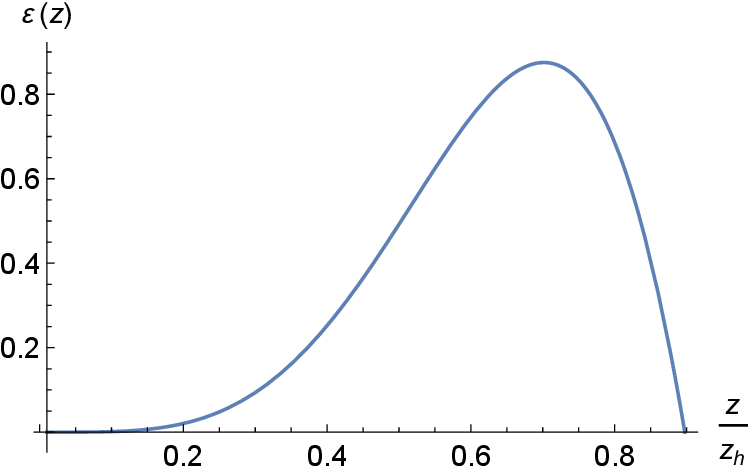}
\end{minipage}
\caption{String profile at some constant speed a) in absence of deformation parameter b) in presence of deformation parameter.}
\label{fig:epsilon}
\end{figure}
Fig.~\ref{fig:epsilon} shows the string ansatz, $\varepsilon(z)$ in a) absence of deformation parameter b) presence of deformation parameter. By solving the integral in right hand side of (\ref{eq:zetaandzetaprim}) numerically, we obtain string profile $\varepsilon(z)$ in these plots.  Contribution of deformation parameter leads to finding of a maximum value of $\varepsilon(z)$. Then, increasing $z$ leads to increasing $\zeta_0(z)$ in plot a) as the string profile is a strictly increasing function of $z$, but in plot b) at nonzero deformation parameter,  $\varepsilon(z)$ has a maximum value.
\newpage
\section{Conclusions}\label{se:con}
We discussed the drag force on  a heavy moving quark in a plasma with a deformed background and deformed probe string. We were motivated by the question arises after applying a deformed AdS: what will happen if deformation parameter appears not only in the background metric but also in the string ansatz? As parameter $c$ deforms the AdS, what are the consequences for probe string? To study this idea we applied a general ansatz for the  probe string  with contribution of deformation parameter. Phenomenologically and according to the drag force on moving heavy quark through a plasma, we tested availability of the ansatz. At the temperature $250$ MeV of QCD, we found the value of deformation parameter as $0.84$ Ge$V^2$  which is in agreement with QCD results. Studying deformed string profile, we showed that while c-independent string ansatz is a strictly increasing function of $z$, c-dependent string ansatz finds a maximum value.\\
\\
\textbf{Acknowledgement}\\
 This work was supported by
Strategic Priority Research Program of Chinese Academy of Sciences (XDB34030301).
ST is supported by the PIFI ( 2021PM0065).

\end{document}